\documentclass{article}
\usepackage[utf8]{inputenc}
\usepackage[yyyymmdd]{datetime}
\usepackage{amssymb}
\usepackage{url}
\usepackage{tabularx}
\usepackage{tikz}
\usepackage{wasysym}
\usepackage{flowchart}
\usepackage{breakurl}
\usepackage{pgf-umlsd}
\usetikzlibrary{shapes,arrows.meta,chains}

\title{A Protocol for Compliant, Obliviously Managed Electronic Transfers}
\author{Geoffrey Goodell\\University College London\\\texttt{g.goodell@ucl.ac.uk}}
\date{\small \textit{This Version: \today}}

\definecolor{ourcol}{rgb}{0.7, 0.2, 0.5}
\definecolor{mfill}{rgb}{0.99, 0.9, 0.6}
\definecolor{mborder}{rgb}{0.8, 0.6, 0.4}
\definecolor{col1}{rgb}{0.7, 0.2, 0.5}
\newcommand{\cza}[1]{{\color{col1}{#1}}}
\definecolor{col2}{rgb}{0.5, 0.2, 0.9}
\newcommand{\czb}[1]{{\color{col2}{#1}}}
\definecolor{col3}{rgb}{0.5, 0.5, 0.5}
\newcommand{\czc}[1]{{\color{col3}{#1}}}
\definecolor{col4}{rgb}{0.0, 0.5, 0.7}
\newcommand{\czd}[1]{{\color{col4}{#1}}}

\newcommand{\cz}[1]{\textit{\textbf{#1}}}

\newcommand*{\mybox}[1]{%
  \fcolorbox{mborder}{mfill}{\raisebox{0pt}[0.5\baselineskip][0.05\baselineskip]{%
    #1}}}
\newcolumntype{L}[1]{>{\raggedright\arraybackslash}p{#1}}
\newcolumntype{C}[1]{>{\centering\arraybackslash}p{#1}}
\newcolumntype{R}[1]{>{\raggedleft\arraybackslash}p{#1}}

\newcommand{\ts}{
    \tikzset{>={Latex[width=3mm,length=3mm]}}
    \tikzstyle{line} = [draw, ->, >=latex, ultra thick]
    \tikzstyle{box} = [rectangle, align=center, text centered]
    \tikzstyle{noshape} = [align=center, text centered]
}

\begin{document}

\maketitle

\begin{abstract}

We describe a protocol for creating, updating, and transferring digital assets
securely, with strong privacy and self-custody features for the initial owner
based upon the earlier work of Goodell, Toliver, and Nakib.  The architecture
comprises three components: a mechanism to unlink counterparties in the
transaction channel, a mechanism for oblivious transactions, and a mechanism to
prevent service providers from equivocating.  We present an approach for the
implementation of these components.

\end{abstract}

\section{Introduction}

We elaborate and specify a protocol for digital payments based upon the work of
Goodell, Toliver, and Nakib~\cite{goodell2022}.  The protocol can be used to
support and implement a range of fungible digital assets, including, without
limitation, cryptocurrency, stablecoins, reserve-backed tokens, and central
bank digital currency (CBDC).  Our proposal is designed to enable the
separation of transactions from the identity of their payers and also to allow
payers to hold fungible assets outside of custodial relationships (accounts).
Our proposal also enables transactions with zero marginal cost to core network
operators (specifically, avoiding native digital tokens or ``gas'' fees) and
avoids requiring issuers to track the ownership of assets in-flight.

\subsection{Our approach to digital currency architecture}

Consider a digital payment system with the following components, with the
objective of combining security and efficiency with self-custody and
irrevocable anonymity for payers in the transaction channel:

\begin{itemize}

\vspace{0.5em}\item \cz{Unforgeable, stateful, oblivious} (\cz{USO})
\cz{assets}~\cite{goodell2022a}, for scalability

\begin{itemize}

\item The issuer does not maintain a database of assets or states (in contrast
to UTXO or state-transition approaches)

\item Notarisation services and ledger operators are \mybox{oblivious} to the
\cz{content} of transactions and do not \cz{process} them

\item The issuer has no role in facilitating consumer transactions

\item Assets are stored in \mybox{non-custodial wallets}, \cz{not the ledger}

\end{itemize}

\vspace{0.5em}\item \cz{Privacy-enhancing technology} to unlink payers from
transactions and transactions from each other, offering privacy by design and
verifiable anonymity for payers

\begin{itemize}

\item Can be done with blind signatures or zero-knowledge proofs

\item Payers are anonymous, and recipients are not

\item Similar privacy model to Chaum, Grothoff, M\"oser~\cite{chaum2021} (and \cz{Taler})

\item USO assets make this privacy model practical

\end{itemize}

\vspace{0.5em}\item A \cz{distributed ledger}, for immutability and institutional trust

\begin{itemize}

\item Nodes are operated by \cz{independent} service providers

\item The ledger grows at a \underline{fixed} rate with \cz{zero marginal cost}
per transaction

\end{itemize}

\end{itemize}

\subsection{Oblivious ledgers}

Consider an associative array mapping keys \czd{$k_i$} to values $v_i$, as follows:

\vspace{-0.7em}\begin{equation}
(\czd{k_0} \rightarrow v_0, \czd{k_1} \rightarrow v_1, \cdots, \czd{k_n} \rightarrow v_n)
\end{equation}

\vspace{0.1em} Define a \cz{root} \cza{$G$} as the output of a well-known hash function
$h$ applied to the associative array:

\vspace{-0.7em}\begin{equation}
\cza{G}=h(\czd{k_0} \rightarrow v_0, \czd{k_1} \rightarrow v_1, \cdots, \czd{k_n} \rightarrow v_n)
\end{equation}

\vspace{0.1em} Consider a ledger operator who accepts contributions of key-value
pairs from its clients and incorporates those pairs into an associative array,
updated once per time period.

\vspace{0.5em} An \cz{oblivious ledger} $L$ is a sequence of roots generated by a
ledger operator at discrete time intervals, as follows:

\vspace{-0.7em}\begin{equation}
\cza{G_{L,0}}, \cza{G_{L,1}}, \cdots, \cza{G_{L,i}}
\end{equation}

\vspace{0.1em} The roots can be linked together into a \cz{blockchain} data
structure by furnishing signatures linking each successive root to the previous
one.  With \cz{Merkle tries}, proofs of \underline{inclusion} or
\underline{exclusion} of a transaction can be done efficiently, in $O(lg\,\,n)$
time.

\subsection{Creating assets}

The initial owner of an asset begins by creating a vector $F_0$ containing three
fields, as follows:

\begin{equation}
F_0 \leftarrow \mybox{$(\czb{u_0}, \cza{G_{L,i}}, \czd{k_1})$}
\end{equation}

\begin{itemize}

\item $\czb{u_0}$ : an arbitrary message (may be empty)

\item $\cza{G_{L,i}}$ : a reference to a specific root $i$ of an oblivious ledger $L$

\item \czd{$k_1$} : the public key matching a new, one-time private key $k^*_1$

\end{itemize}

\vspace{1em} The initial owner then creates an asset by combining the vector
$F_0$ with a genesis signature, as follows:

\begin{equation}
A_0 \leftarrow (\mybox{$F_0$}, s(h(\mybox{$F_0$}),\czd{k_0}))
\end{equation}

\begin{itemize}

\item $s(\czc{d}, \czc{k})$ : signed copy of some data $d$ verifiable by a public key $k$

\item \czd{$k_0$} : a long-term key held by some issuer (or the initial owner, if the
owner is allowed to create assets)

\end{itemize}

\subsection{Updating assets}

To update an asset, an owner must \cz{register} the update with a ledger
operator called an \cz{integrity provider} (which is essentially a notarisation
service, for for which we also use the term \cz{relay}).  The owner at sequence
number $j$ must create an \textit{update vector} $F_j$ containing four fields,
as follows:

\begin{equation}
F_j \leftarrow \mybox{$(\czb{u_j}, \cza{G_{L,i}}, \czd{k_{j+1}}, h(A_{j-1}))$}
\end{equation}

\begin{itemize}

\item $\czb{u_j}$ : an indication of the type or nature of the update

\item $\cza{G_{L,i}}$ : a reference to a specific root $i$ of an oblivious
ledger $L$ (or empty, if the current ledger is to be retained)

\item \czd{$k_{j+1}$} : the public key matching a new, one-time private key $k^*_{j+1}$

\item $h(A_{j-1})$ : the hash of the last version of the asset prior to the update

\end{itemize}

\vspace{1em} The asset owner then signs the update vector, creating an update
$U_j$, and the updated asset $A_j$ is defined by concatenating the previous
version of the asset $A_{j-1}$ with the update:

\begin{equation}
U_j \leftarrow (\mybox{$F_j$}, s(h(\mybox{$F_j$}),\czd{k_j}))
\end{equation}
\begin{equation}
A_j \leftarrow A_{j-1} \,\vert\vert\, U_j
\end{equation}

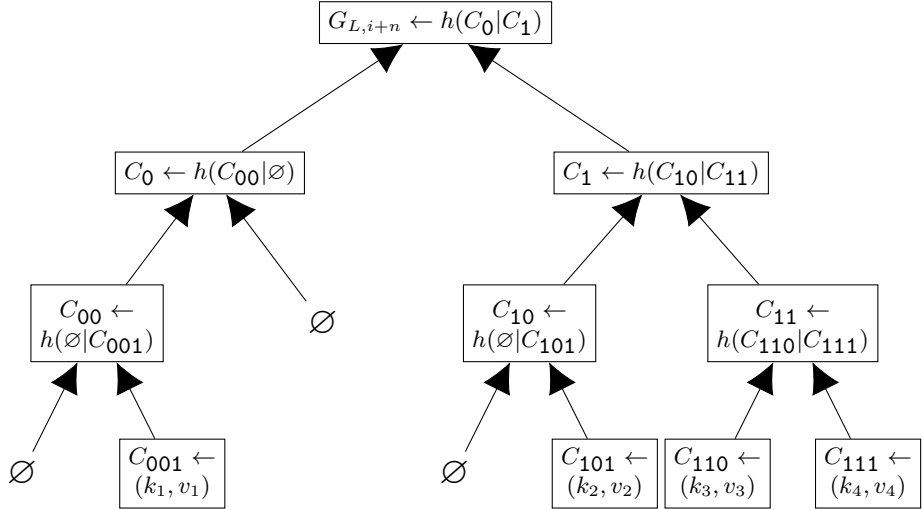
\begin{figure*}[!ht]
\begin{center}
\hspace{-0.8em}\scalebox{1}{
\begin{tikzpicture}[>=latex, node distance=3cm, font={\sf \small}, auto]\ts
\tikzset{>={Latex[width=4mm,length=4mm]}}
\node (box1) at (0, 0) [box, draw] {$G_{L,i+n} \leftarrow h(C_\texttt{0}\,||\,C_\texttt{1})$};
\node (box2) at (-3, -2) [box, draw] {$C_\texttt{0} \leftarrow h(C_\texttt{00}\,||\,\varnothing)$};
\node (box3) at (3, -2) [box, draw] {$C_\texttt{1} \leftarrow h(C_\texttt{10}\,||\,C_\texttt{11})$};
\node (box4) at (-4.5, -4) [box, draw, text height=2em, text centered, align=center] {$C_\texttt{00} \leftarrow $\\$h(\varnothing\,||\,C_\texttt{001})$};
\node (box5) at (-1.5, -4) [box] {\Large $\varnothing$};
\node (box6) at (1.25, -4) [box, draw, text height=2em, text centered, align=center] {$C_\texttt{10} \leftarrow $\\$h(\varnothing\,||\,C_\texttt{101})$};
\node (box7) at (4.75, -4) [box, draw, text height=2em, text centered, align=center] {$C_\texttt{11} \leftarrow $\\$h(C_\texttt{110}\,||\,C_\texttt{111})$};
\node (box8) at (-5.5, -6) [box] {\Large $\varnothing$};
\node (box9) at (-3.5, -6) [box, draw] {$C_\texttt{001}\leftarrow$\\$(k_1,v_1)$};
\node (box10) at (0.25, -6) [box] {\Large $\varnothing$};
\node (box11) at (2.25, -6) [box, draw] {$C_\texttt{101}\leftarrow$\\$(k_2,v_2)$};
\node (box12) at (3.75, -6) [box, draw] {$C_\texttt{110}\leftarrow$\\$(k_3,v_3)$};
\node (box13) at (5.75, -6) [box, draw] {$C_\texttt{111}\leftarrow$\\$(k_4,v_4)$};

\draw[->] (box2) -- (box1);
\draw[->] (box3) -- (box1);
\draw[->] (box4) -- (box2);
\draw[->] (box5) -- (box2);
\draw[->] (box6) -- (box3);
\draw[->] (box7) -- (box3);
\draw[->] (box8) -- (box4);
\draw[->] (box9) -- (box4);
\draw[->] (box10) -- (box6);
\draw[->] (box11) -- (box6);
\draw[->] (box12) -- (box7);
\draw[->] (box13) -- (box7);
\end{tikzpicture}}
\end{center}

\caption{A schematic representation of a Merkle trie with root $G_{L,i+n}$}.

\label{f:proof-of-inclusion}
\end{figure*}

The asset owner submits \czd{$k_j$} and $s(h($\mybox{$F_j$}$), \czd{k_j})$ to
the integrity provider specified in $F_{j-1}$ in exchange for a \cz{proof of
inclusion} $p(G_{L,i+n},k_j,F_j)$, where $n$ represents the number of roots of
$G_L$ that have been sequentially produced by the ledger operator since the
creation of $G_{L,i}$.  A proof of inclusion demonstrates that a key-value pair
$(k_j,F_j)$ has been inserted into an associative array with root $G_{L,i+n}$.
This can be done efficiently with a Merkle trie, a data structure that encodes
the keys of an associative array into a tree structure; see
Figure~\ref{f:proof-of-inclusion}.

The proof of inclusion of a value $C_\texttt{110}$ corresponding to a key $k_3$
whose first three bits are \texttt{110} would be given as
$(k_3,C_\texttt{0},C_\texttt{10},C_\texttt{111})$.  The proof of inclusion
provides all of the information necessary to confirm that the pair $(k_3,v_3)$
is included in the Merkle trie, since $G_{L,i+n}$ can be quickly calculated to
equal $h(C_\texttt{0}\,||\,h(C_\texttt{10}\,||\,h(C_\texttt{110}\,||\,C_\texttt{111})))$.  It
is also possible to provide a \cz{proof of exclusion} to demonstrate the
non-existence of an entry with some key $k$ by showing a proof of inclusion for
some pair with $k'\neq k$ that would appear in the same location or along the
path to a pair with key $k$ if it were included.

Finally, the \cz{proof of provenance} $P_j$ of an asset $A_j$ comprises the proof
of provenance for $P_{j-1}$ plus the new proof of inclusion:

\begin{equation}
P_j \leftarrow P_{j-1} \,\vert\vert\, p(G_{L,i+n},k_j,F_j)
\end{equation}

The proof of provenance $P_0$ for the newly created asset $A_0$ is the empty
set, $\varnothing$.

\subsection{Transferring assets}

If an owner at sequence number $j$ (the ``old owner'') wishes to transfer an
asset to a new owner (at sequence number $j+1$), then the owner of sequence
number $j$ must create an update vector:

\begin{equation}
F_j \leftarrow \mybox{$(\czb{u_j}, \cza{G_{L,i}}, \czd{k_{j+1}}, h(A_{j-1}))$}
\end{equation}

To complete the transfer, \czd{$k_{j+1}$} \underline{must} be provided by the
new owner.  Then, the old owner creates the update:

\begin{equation}
U_j \leftarrow (\mybox{$F_j$}, s(h(\mybox{$F_j$}),\czd{k_j}))
\end{equation}
\begin{equation}
A_j \leftarrow A_{j-1} \,\vert\vert\, U_j
\end{equation}

\vspace{1em} Once the old owner shares $A_j$ and $P_j$ with the new owner (see
above), the new owner has \cz{possession} of the asset.

\vspace{1em} Once \czd{$k_j$} and $s(h($\mybox{$F_j$}$), \czd{k_j})$ are shared
with the integrity provider specified in $F_{j-1}$, the new owner has
\cz{control} of the asset.  (It is possible for the new owner to have control
without possession.)

\subsection{A Chaumian Mint (or ZKP) for Privacy}

Following Chaum~\cite{chaum1982}, suppose that we have a function $b$ such
that:

\vspace{-0.5em}\begin{equation}
b^{-1}(s(b(d), k))=s(d, k)
\end{equation}

Then, the initial owner of an asset can send $b(h(\mybox{$F_0$}))$ to an
issuer, who will be able to create the signature $s(b(h(\mybox{$F_0$})),
\czd{k_{0}})$.  The initial owner can then apply $b^{-1}$, yielding
$s(h(\mybox{$F_0$}), \czd{k_{0}})$.

\vspace{0.5em} If the asset $A_0$ represents something fungible, then the
initial owner can transfer this token \textit{without revealing its identity}
or any pseudonym.

\vspace{0.5em} \cz{Alternatively}, the initial owner can furnish to the new
owner a \cz{zero-knowledge proof}~\cite{friolo2025} linking the vector
\mybox{$F_0$} to the ledger \cza{$G_L$}.

\subsection{Using DLT to prevent equivocation}

There is a persistent risk that notarisation services might \cz{equivocate} by
producing two different versions of some \cza{$G_{L,i}$}.

\vspace{1em} To address this risk, integrity providers \cz{may} periodically
commit their roots \cza{$G_{L,i}$} to a distributed ledger, which can combine
the roots produced by $n$ different ledgers at a given time $t$, as follows:

\begin{equation}
\cza{G_{D,t}}=h(\czd{k_{L_1}} \rightarrow \cza{G_{L_1,t}}, \czd{k_{L_2}} \rightarrow \cza{G_{L_2,t}}, \cdots, \czd{k_{L_n}} \rightarrow \cza{G_{L_n,t}})
\end{equation}

\vspace{0.5em} The sequence of roots $G_{D,0}, \cdots, G_{D_t}$ can be linked
together into a blockchain data structure, using the same approach taken by
individual integrity providers to manage their ledgers.  Proofs of
\underline{inclusion} or \underline{exclusion} of integrity provider roots can
be furnished by participants in the DLT system, and they can be combined with
proofs provided by integrity providers to provide assurance that the integrity
providers did not equivocate.  The choice to use a DLT is \cz{optional} and
ultimately for the issuer to decide.

\vspace{1em} In general, proofs can be \cz{stacked}: roots of integrity
providers or DLT systems can be committed into ledgers successively, without
limit.

\begin{figure*}[ht]
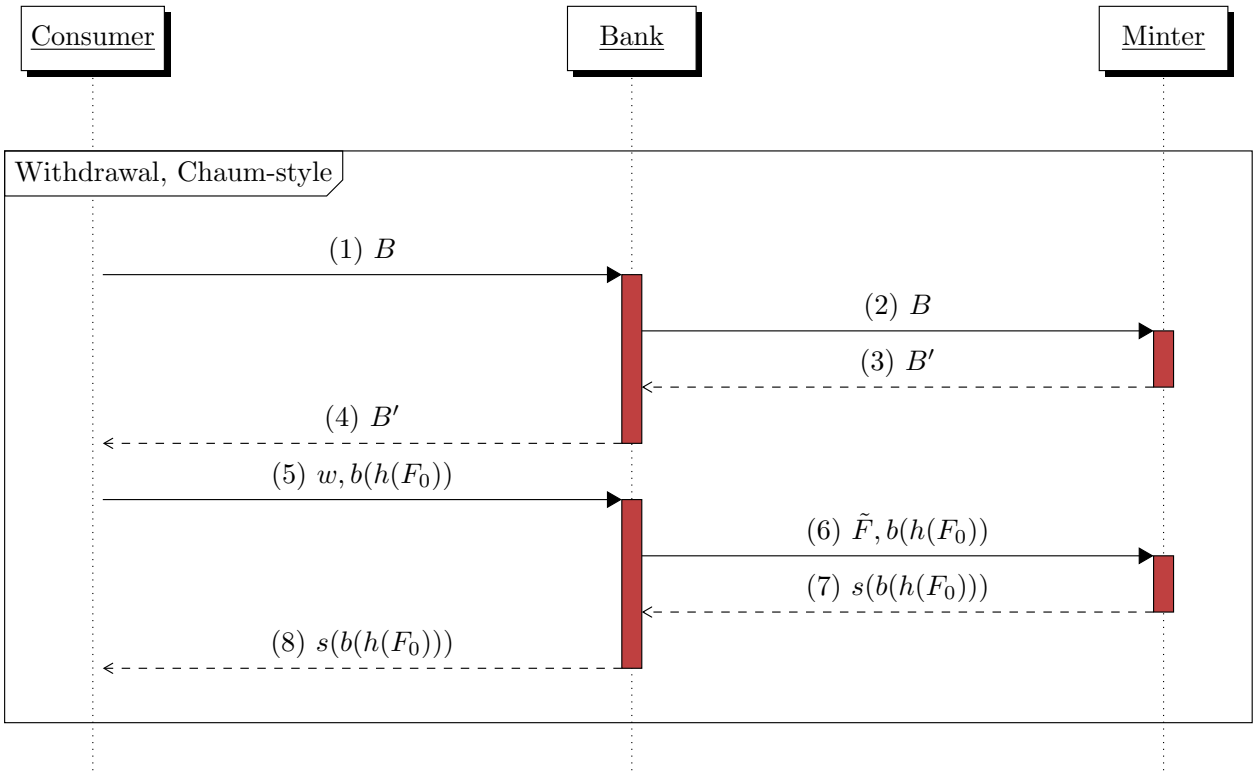

  \centering
  \resizebox{0.99\textwidth}{!}{
  \begin{sequencediagram}
    \tikzset{threadstyle/.style={fill=red!50!gray}}
    \renewcommand\unitfactor{0.7}
    \newinst[0]{consumer}{Consumer}{}
    \newinst[5]{bank}{Bank}{}
    \newinst[5]{minter}{Minter}{}

    \begin{sdblock}{Withdrawal, Chaum-style}{}
    \path (0,0) -- (15,0);
        \begin{call}{consumer}{
            (1) $B$
        }{bank}{
            (4) $B'$
        }
        \begin{call}{bank}{
            (2) $B$
        }{minter}{
            (3) $B'$
        }
        \end{call}
        \end{call}
        \begin{call}{consumer}{
            (5) $w,b(h(F_0))$
        }{bank}{
            (8) $s(b(h(F_0)), k_0)$
        }
        \begin{call}{bank}{
            (6) $\tilde{F},b(h(F_0))$
        }{minter}{
            (7) $s(b(h(F_0), k_0))$
        }
        \end{call}
        \end{call}
    \end{sdblock}

  \end{sequencediagram}}
  \caption{UML sequence diagram for withdrawing an asset using the Chaum method.}
  \label{f:withdrawal-chaum}
\end{figure*}

\begin{figure*}[ht]
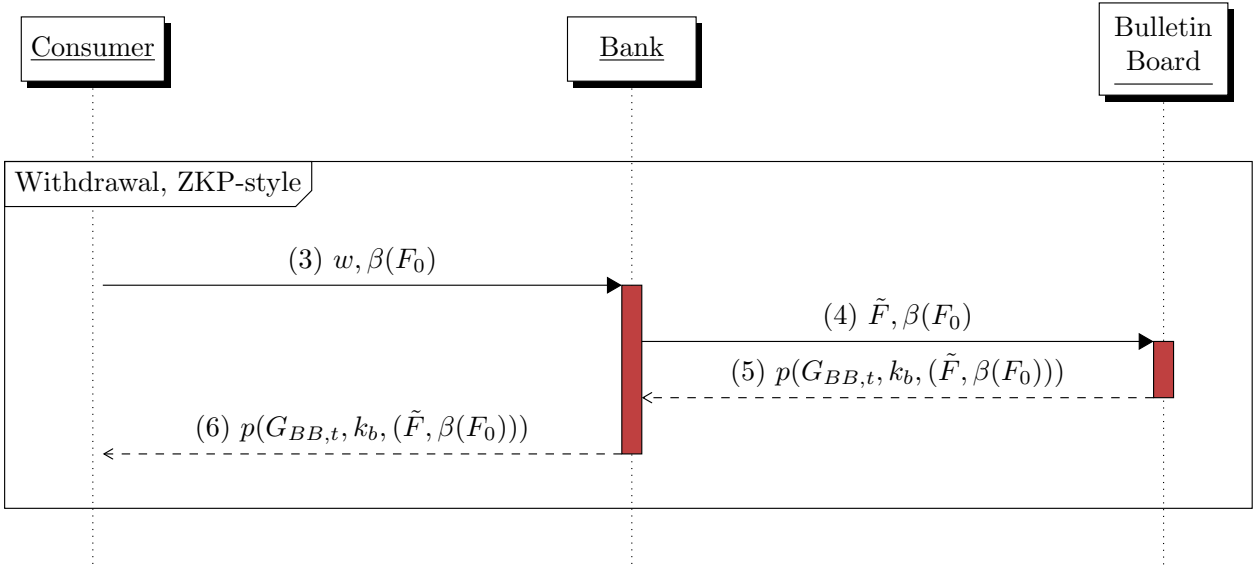

  \centering
  \resizebox{0.99\textwidth}{!}{
  \begin{sequencediagram}
    \tikzset{threadstyle/.style={fill=red!50!gray}}
    \renewcommand\unitfactor{0.7}
    \newinst[0]{consumer}{Consumer}{}
    \newinst[5]{bank}{Bank}{}
    \newinst[5]{bb}{\begin{tabular}[t]{@{}c@{}}Bulletin\\Board\end{tabular}}{}

    \begin{sdblock}{Withdrawal, ZKP-style}{}
    \path (0,0) -- (15,0);
        \begin{call}{consumer}{
            (3) $w, \beta(F_0)$
        }{bank}{
            (6) $p(G_{BB,t},k_b,(\tilde{F}, \beta(F_0)))$
        }
        \begin{call}{bank}{
            (4) $\tilde{F}, \beta(F_0)$
        }{bb}{
            (5) $p(G_{BB,t},k_b,(\tilde{F}, \beta(F_0)))$
        }
        \end{call}
        \end{call}
    \end{sdblock}

  \end{sequencediagram}}
  \caption{UML sequence diagram for withdrawing an asset using the ZKP method.}
  \label{f:withdrawal-zkp}
\end{figure*}

\section{Protocol specification}

\begin{figure*}[!ht]
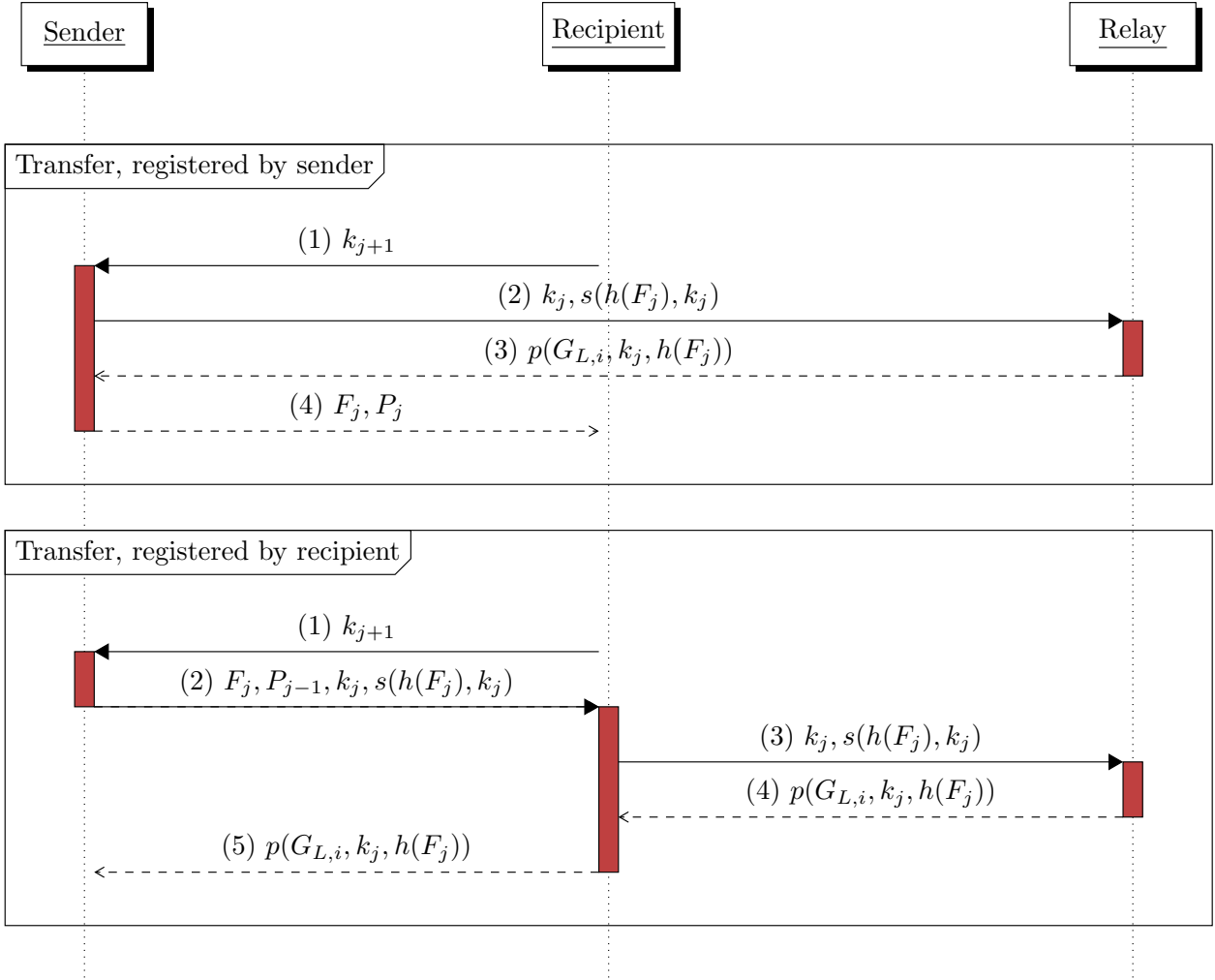

  \centering
  \resizebox{0.99\textwidth}{!}{
  \begin{sequencediagram}
    \tikzset{threadstyle/.style={fill=red!50!gray}}
    \renewcommand\unitfactor{0.7}
    \newinst[0]{sender}{Sender}{}
    \newinst[5]{recipient}{Recipient}{}
    \newinst[5]{relay}{Relay}{}

    \begin{sdblock}{Transfer, registered by sender}{}
    \path (0,0) -- (14.8,0);
        \begin{call}{recipient}{
            (1) $k_{j+1}$
        }{sender}{
            (4) $F_j,P_j$
        }
        \begin{call}{sender}{
            (2) $k_j,s(h(F_j),k_j)$
        }{relay}{
            (3) $p(G_{L,i},k_j,h(F_j))$
        }
        \end{call}
        \end{call}
    \end{sdblock}

    \begin{sdblock}{Transfer, registered by recipient}{}
    \path (0,0) -- (14.8,0);
        \begin{call}{recipient}{
            (1) $k_{j+1}$
        }{sender}{}
        \end{call}
        \addtocounter{seqlevel}{-1}
        \begin{call}{sender}{
            (2) $F_j,P_{j-1},k_j,s(h(F_j),k_j)$
        }{recipient}{
            (5) $p(G_{L,i},k_j,h(F_j))$
        }
        \begin{call}{recipient}{
            (3) $k_j,s(h(F_j),k_j)$
        }{relay}{
            (4) $p(G_{L,i},k_j,h(F_j))$
        }
        \end{call}
        \end{call}
    \end{sdblock}
  \end{sequencediagram}}
  \caption{UML sequence diagram for transferring an asset; the transfer can be registered by either the sender or the recipient.  If the sender used the ZKP method to withdraw the asset, then the sender can transfer a zero-knowledge proof $\pi$ relating $F_0$ to ($\tilde{F},\beta(F_0))$ to the recipient along with $F_j$ so that the recipient can verify that asset was created via a valid burning operation.}
  \label{f:transfer}
\end{figure*}

For our protocol sequence diagrams (see Figures~\ref{f:withdrawal-chaum},
\ref{f:withdrawal-zkp}, and~\ref{f:transfer}), we use the following symbols:

\begin{itemize}

\item $w$ : withdrawal authorisation for $q$ units of currency.

\item $B$ : request to initialise the blinding protocol.

\item $B'$ : initialisation of the blinding protocol.

\item $\tilde{F}$ : voucher or tokens for redemption (tokens may be spent tokens).

\item $p(G_{L,i},k_j,F_j)$ : proof of inclusion indicating the binding of key $k_j$ to
update $F_j$ within a tree with root $G_{L_i}$.

\item $\beta(F_0)$ : a signed ZK commitment to a new asset $F_0$ by an actor
who will spend $F_0$ in the future.

\end{itemize}

\section*{Acknowledgements}

We thank Dann Toliver of TODAQ Financial for the proposed name for the protocol
and Jason Polis for his methodical and insightful review of the design.  We
thank Professor Tomaso Aste for his continued support for our work on digital
currencies and digital payment systems, and we also acknowledge the Future of
Money Initiative at University College London and the Systemic Risk Centre at
the London School of Economics.

\end{document}